# DEVELOPMENT OF APPLICATION FOR DISCOVERING AND BINDING TO PUBLISHED GEOSPATIAL PROCESSES IN DISTRIBUTED ENVIRONMENTS

**Evgeny V. Shulkin** – Pacific Geographical Institute, Far Eastern Branch of the Russian Academy of Sciences, Vladivostok, **Russian Federation.**

**Sergey M. Krasnopeyev** – Pacific Geographical Institute, Far Eastern Branch of the Russian Academy of Sciences, Vladivostok, **Russian Federation.**

## ABSTRACT

Nowadays, society has recognized that the lack of access to spatial data and tools for their analysis is the limiting factor of economic development. It came to the realization that without the single information space, which is implemented in the form of spatial data infrastructures, a progressive business development is impossible.

Spatial data infrastructures will support a variety of tasks, which requires the binding of geospatial information from multiple sources.

In the last few years, the rate of progress in spatial data collection was higher, than in management and analysis of data. Infrastructures allow the accumulated data to be available to large groups of users, and infrastructure of analysis allows the data to be effectively used for such tasks as municipal planning, science research, etc. Moreover, free access to the information resources and instruments of analysis will serve as an additional impulse to development of application models in corresponding areas of expertise.

The goal of this paper is to indicate possible solutions to the client-side problems of spatial data analysis in distributed environments, using the developing application for data analysis as an example.

**Keywords:** spatial data analysis, Open Geospatial Consortium, spatial data infrastructure, Web Processing Service, client for spatial data analysis.

## OGC STANDARDS

The problem of geospatial interoperability is the key factor in deployment infrastructures for publishing and spatial data analysis. Open Geospatial Consortium (OGC) [1] elaborates the approaches to solve this problem.

By OGC, web services are technological basis for interoperability. Consortium of more than 144 independent companies suggested a number of Web standards to simplify publishing of geospatial data in distributed environments.

The most important and popular are the standards that describe output protocol and issue of publication of spatial data over the Internet. These standards include **Web Map Service** (WMS), **Web Feature Service** (WFS) and **Web Coverage Service** (WCS) [2].

## STANDARD FOR SPATIAL DATA ANALYSIS

According to OGC specification, **Web Processing Service** (WPS) [2] is a standard for geospatial data analysis. It provides services for geoprocessing raster and vector spatial data. Operations can be simple (eq, geometric operations on vector data), and very complex, up to the calculation of global environments models. Standard defines data transfer protocol, format of the commands for process execution and data retrieving. This is a service publication in its purest form.

Standard defines three main operations:
- **GetCapabilities** for extracting web services metadata and list of the processes.
- **DescribeProcess** returns a detailed description of the selected process, including information about the input and output parameters.
- **Execute** starts the process and returns result.

Web Processing Service accepts these commands in XML format or as a URL-encoded request. The standard is very flexible, it is ensured by its properties:
- Input and output data could be either references to the data, or be integrated into the body of the request or response.
- If the answer is simple, such as an image in the GIF format, the web service can return it directly, without wrapping in XML.
- Web service supports various formats of input and output data. For example, numbers and strings, bounding boxes, vector data in GML format, and binary data. Theoretically, the list can go on, as web service includes the opportunity to supplement its own data format.

**WPS STANDARD IMPLEMENTATION**

In a general sense, implementation of the WPS standard involves the deployment of two components:
- The server side, implemented as a server application. Most of the currently existing implementations written in Java, but there are servers that are written in C and Python.
- The client side application. Could be both desktop and browser.

**SERVER SIDE APPLICATION**

At first, we will briefly touch the issue of server side application. Server is the main component in WPS deployment. It contains a list of processes, accepts requests and retrieves data from them, passes data to the process and generates response with the result. When one talks about the implementation of the WPS standard, it means the server side.

Currently, there are several successfully developing open-source projects, that implements server side WPS. For example, deegree WPS [3], ZOO Project [4], GeoServer WPS [5] and 52° North WPS [6].

**CLIENT SIDE APPLICATION**

Server side alone could not provide all the requirements arising from the complex analysis of spatial data. Of course, using the URL-encoded requests through browser address bar, it is possible to directly contact web service, and even get some useful information. For example, metadata and the list of processes. But here we must consider, that information will be in a raw XML, which is pretty hard to understand and even harder to use manually in next steps of data analysis.

It is therefore necessary to deploy the second component of the infrastructure of data analysis – the client application.

**REQUIREMENTS FOR CLIENT SIDE**

The base algorithm for client application is determined by WPS logic, i.e. client-side interaction may consist of three main steps. Each step outputs the data from previous step and ends with sending the corresponding request to service.

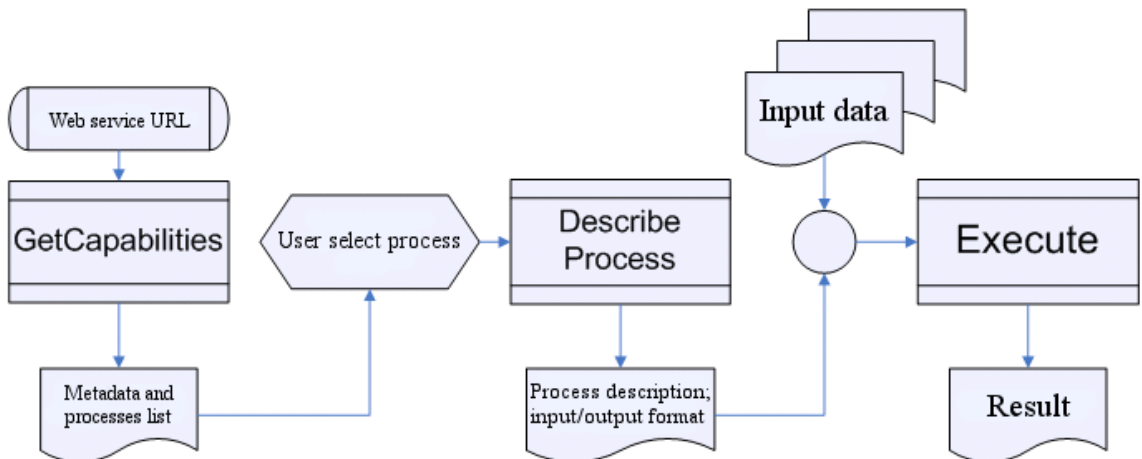

Img. 1. Stages of web processing by WPS standard

Here we closely look at each step and consider the requirements for interface and logic, which should be met by client application on corresponding step.

**GetCapabilities**

At this point, client needs to extract metadata from web service. User specifies only web service URL and nothing more, so the interface at this stage is pretty simple:

- **Text field** to manually input URL and / or **drop-down list** with some default addresses.

By parsing response from **GetCapabilities** request, client should compose necessary data structures for next stages:

- Web service title and abstract.
- Supported WPS version and operations.
- List of available processes.

**DescribeProcess**

From the client point of view, this stage serves for selecting necessary process from the list of available processes.

User must select only one process from the list. Also, we need to display some additional information from previous stage. Therefore, interface should consist of:

- **Labels** for web services title, abstract, version, total number of processes, etc.
- **Drop-down list** with titles and / or abstracts of available processes.

The result of this stage is the complete information about selected process:

- Title and abstract.

- Lists of input and output parameters, defined by title, type and format.

**Execute**

The last and the most complex stage, devoted to the definition of input parameters. As usual, we need some **labels** to display processes metadata. Then it became complicated, because we need to display input elements for each input parameter, depending on its type.

According to WPS standard, there are three types of data, which can be used as input and output parameters – **Literal**, **Complex** and **Bounding Box**.

Let's look at each case independently:
- Literal input data:
  Literal data means strings, numbers, etc. In the trivial case, we only need one **text field** for user to enter the number. But it can become complicated, if process, by some chance, requires sophisticated literal data like arrays, tables, etc.
- Complex and BBox input data:
  Binary and XML complex data is used to transfer spatial data. Vector data is usually transferred in GML format. There are several ways to deal with such situation:
    1. We can extract vector spatial data from **map** and convert it to **GML format**. Communication between the client and the map can be performed by API. In this case, the map should support **vector layers** and an **API for data export**.
    2. Or, we can request XML data from **remote web service** which supports **Web Feature Service** OGC standard. By WFS standard, spatial data exports as feature collection, i.e. array of geometries with some attribute information. First of all, we need input elements to compose WFS request:
       - **Text field** and / or **drop-down list** for service URL.
       - **Drop-down list** for the list of available layers.
       - **Editable grid** for some standard WFS filters, including maximum number of features, feature id, attribute information, etc.
  
  We can send composed query to WPS as **reference** input data, so the web service itself will request the data. Or, we can **request the data on client side** and send it to WPS. This can be useful, when we need to process the spatial data before sending it to WPS. For example, extract geometry from feature collection, because some processes can handle only geometries, without attributes. Of course, we need some **toggle** to switch between these options.

Given this data, we can compose the request to web service and execute it. The result will be the XML response from service, which we need to parse and output.

**Process the results**

On this additional stage user is provided a number of options to deal with the results. A list of outputs is given, along with the interface elements. Let's review the possible options:
- Literal output data:
  This is the simplest case. User can only view the literal data via **label**, or **export** it to text format.

- Complex and BBox output data:
  In this case, the number of options is also limited. User can **draw** the result on the maps vector layer. Or user can **save** spatial data to some format to use later.

**IMPLEMENTATION OF THE CLIENT APPLICATION**

There are commercial software packages for professional and demanding tasks of spatial data analysis. For example, ArcGIS series by ESRI. But the main objective of infrastructures of data analysis is to provide free access to spatial data and analytical tools to the largest possible number of users.

Such an approach is justified in the educational process, as well as decision-making systems. In this case, the speed of access to infrastructures resources and ease of use are critical. The problem of cross-platform arises.

Therefore, during the development of client application for data analysis, several design decisions were made:
- The easiest way to solve the compatibility problem on different platforms is to implement client as a **browser-based application** or website. Difficult cross-platform problem is replaced by trivial cross-browser problem. In addition, the browser-based implementation solves the problem of mass access.
- The next step in the development of browser-based client is the idea to use the power of already established mapping sites and geo-portals. Client application in this case should be implemented as a **plug-in software module**, which has a standardized interface for receiving input data from third-party application.
- Previous point, because of the huge variety of mapping sites and desktop application, is unnecessarily complex and somewhat utopian. More realistic would be to create a module based on the existing environment for the development of mapping services. For example, a **popular cartographic library OpenLayers** [7], which, to date, is one of the de facto standards for creating browser-based mapping client side. Thus we lost the clients versatility, but obtain the ease of use and connection.
- From the developer's perspective, the client's **modules should be independent of each other**. For example, the individual stages of interaction with the WPS should be allocated to different classes. In this case, there is no need to redo all the logic of the client, if the logic of one the stages will change.
- The preceding point will be very well complemented by the ideology of **open source software**. Thus, even if we did not consider anything in the development of the basic version of the client, third-party developer can always adapt it to fit their needs, writing their own desired set of functions and interfaces for selecting data.
- It also seems a logical step to create a **separate web service for processing requests and responses** from the data analysis service, i.e. the bridge between the client and WPS. The stages for receiving metadata and processes descriptions are pretty much standardized for all classes of processing problems, so the existence of such web service will greatly facilitate client's logic and allow the developer to focus on the problem of data input.

In line with these proposals, we created a universal client module for analysis of spatial data by WPS standards. At this point, the client module is cross-platform, browser-based and supports integration with mapping applications, based on OpenLayers.

Supports:
- Obtaining web service metadata.
- Processing of vector spatial data located on a layer of OpenLayers.
- Processing of vector spatial data from remote web service by WFS standard. It can handle both the geometry and collection of features.
- Display results as vector data on a layer of OpenLayers.

System requirements:
- Mapping client must support OpenLayers.
- The client also must support the management of vector layers in OpenLayers. For example, WFS layers or drawing graphic primitives.

As a test mapping site, a geo-portal, developed in Pacific Geographical Institute is used. Geo-portal is based on OpenLayers and supports creation of vector graphics [8].

Technically, the client module is a set of scripts on JavaScript and PHP, as well as style sheets. The module is loaded via AJAX request and displayed in a popup window with the help of the third-party library.

**DEMONSTRATION OF THE CLIENT MODULE**

Let's demonstrate the work of the client module on a typical problem. Consider a simple task – to build an exclusion zone around the site of the road. Such problems often arise in the design of urban projects.

**Step №1**

Load roadmap in the mapping client. Outline the pavement, using a draw line tool. The problem reduces to the construction of a **buffer zone** around a selected section of the road.

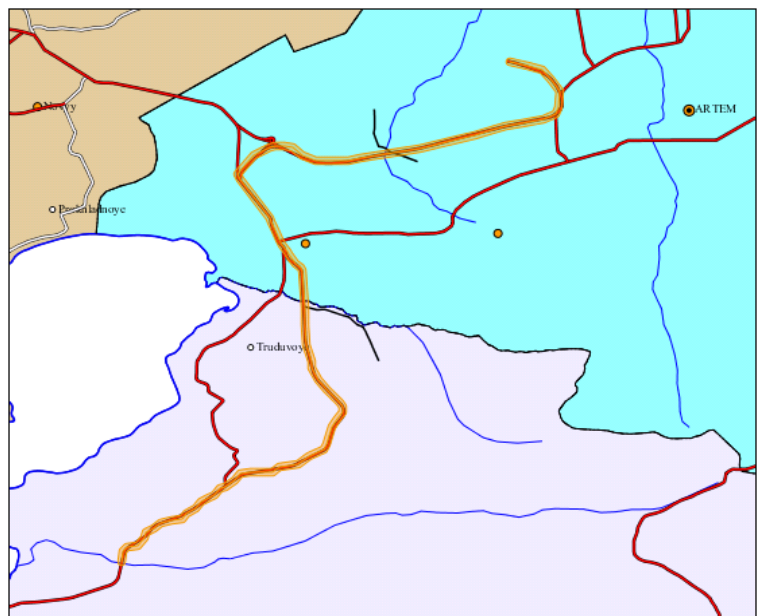

Img. 2. Baseline vector data

**Step №2**

Via the client module, we choose a web service for data analysis and ask for its metadata. After receiving the metadata, we can be sure that the web service supports the desired operation. Choose from the list.

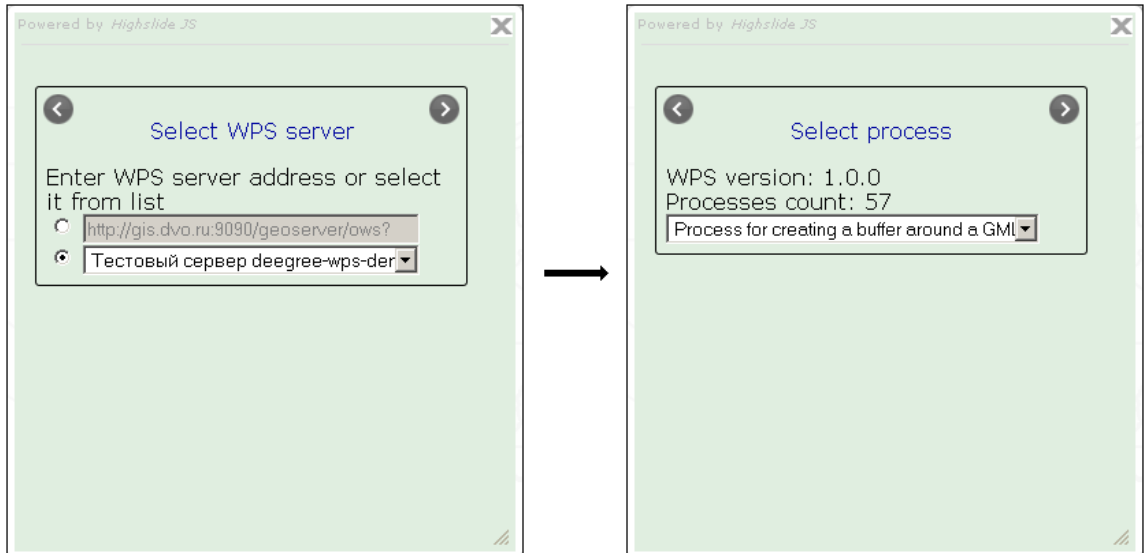

Img. 3. Interface for selecting web service and process

**Step №3**

After that we need to define the input parameters to run the procedure. The interface allows you to select an element of the vector layer and to determine the numerical parameters, in this case, the width of the buffer zone.

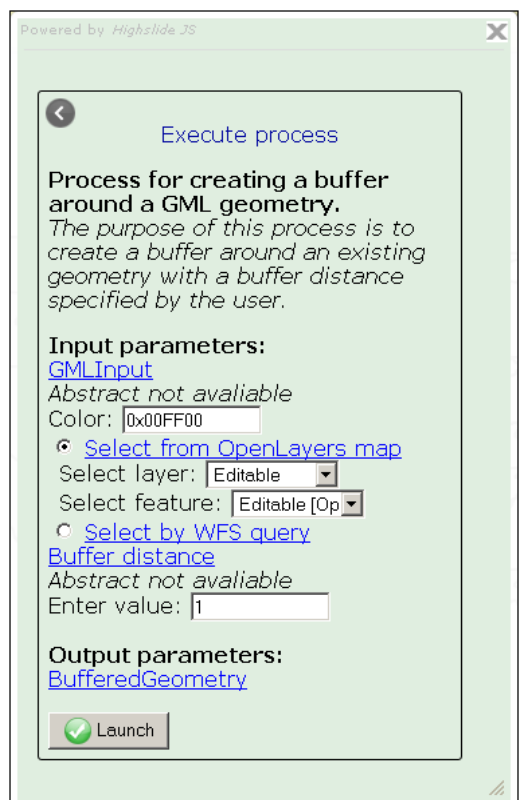

Img. 4. Interface for selecting input values

**Step №4**

Result of the operation, a buffer zone, is displayed on a new vector layer.

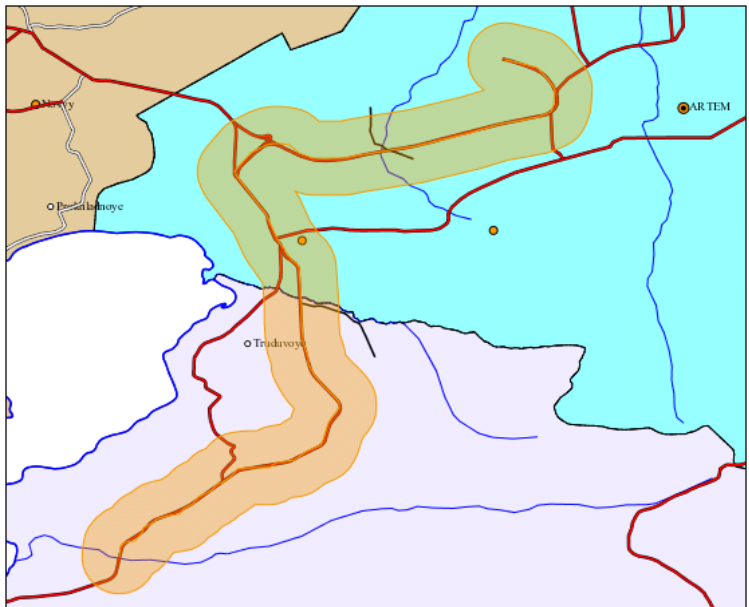

Img. 5. Result of the operation

**CONCLUSION**

Slow development in this area is understandable, since really serious problems in the spatial data analysis are successfully resolved via desktop business applications. But do not forget that the use of open infrastructures provides its own advantages, including mass and openness. That is why this area should develop as much as possible. The presence of a universal client for spatial data analysis or recommendations for its creation will be small, but useful achievement.